\documentclass[english]{article}
\usepackage{graphicx,psfrag}
\usepackage{cite}
\def\no{\noindent}
\def\bc{\begin{center}}
\def\ec{\end{center}}
\def\vs{\vskip0.5cm}
\def\beq{\begin{equation}}
\def\eeq{\end{equation}}

\begin{document}


\title{Short note on the excitonic Mott phase \\
}
\author{Klaus Ziegler$^{1}$, Oleg L. Berman$^{2,3}$ and Roman Ya. Kezerashvili$^{2,3}$, 
\\
$^1$   Institut f\"ur Physik, Universit\"at Augsburg\\
D-86135 Augsburg, Germany\\ 
$^{2}$Physics Department, New York City College of Technology, The City University of New York, \\
Brooklyn, NY 11201, USA \\
$^{3}$The Graduate School and University Center, The City University of New York, \\
New York, NY 10016, USA \\
}
\date{\today}

\maketitle

\no
Abstract:

\no
An exciton gas on a lattice is analyzed in terms of a convergent hopping expansion. For a given
chemical potential our calculation provides a sufficient condition for the hopping rate
to obtain an exponential decay of the exciton correlation function. This result indicates the existence
of a Mott phase in which strong fluctuations destroy the long range correlations in the exciton gas
at any temperature, either by thermal or by quantum fluctuations.


\section{Introduction}

\begin{figure*}[t]
\begin{center}
\vskip-2.5cm
\includegraphics[width=11cm]{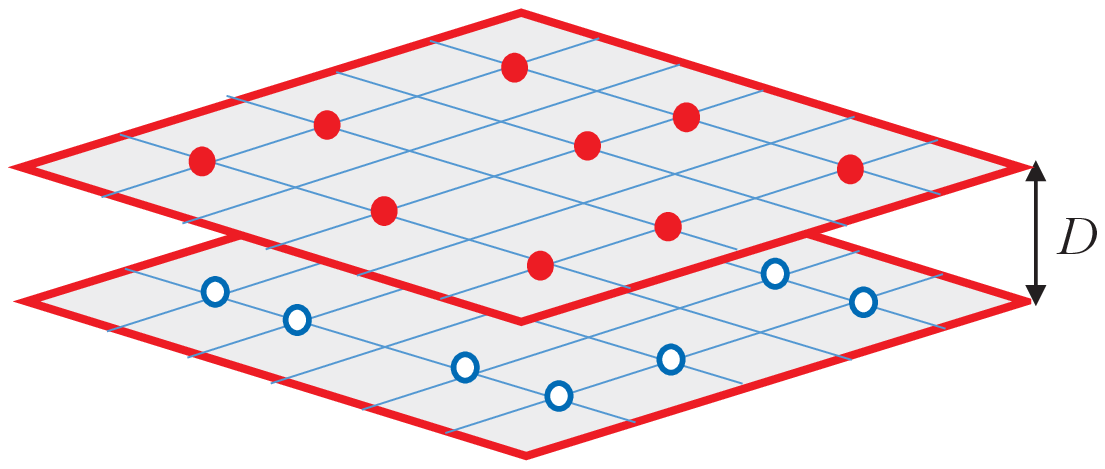}
\vskip-6.5cm
\caption{
Two coupled quantum wells with lattice gates, realized for electrons (filled circles) in one well and holes (empty circles) in the other.
}
\label{fig:system}
\end{center}
\end{figure*}

Coupled quantum wells represents a class of systems which allows us to study strongly interacting
particles under controllable conditions. They are conceptually simple: negative
electrons are trapped in a two-dimensional plane, while an equal
number of positive holes is trapped in a parallel plane a distance
$D$ away (see Fig. \ref{fig:system}). One of the appeals of such systems is that the electron and
hole wavefunctions have very little overlap, so that the excitons
can have very long lifetime  ($> 100$ ns), and therefore, they can be
treated as metastable particles to which quasi-equilibrium statistics applies. 

The exciton gas is effectively a hardcore Bose gas, whose groundstate should be a superfluid at low density 
or a Mott state at high density. The latter requires a lattice such that a lattice commensurate state can
be formed.
A superfluid state of excitons in coupled quantum wells was predicted some time ago in
Ref.~\cite{Lozovik}. Several subsequent theoretical studies 
\cite{Shevchenko,Lerner,Dzjubenko,Kallin,Yoshioka,Littlewood,Vignale,Ulloa} 
have suggested that superfluidity should be manifested as persistent electric currents, 
quasi-Josephson phenomena and unusual properties in strong magnetic fields. 
In the past ten years, a number of experimental studies have focused on the observation of the superfluid
behavior \cite{Snoke_paper,Snoke_paper_Sc,Chemla,Krivolapchuk,Timofeev,Zrenner,Sivan,Snoke}.
The transition from an exciton gas to an electron plasma in GaAs–GaAlAs quantum wells was analyzed in the framework of many body 
effects, considering the dynamical screening of the Coulomb interaction in the one-particle properties of 
the carriers and in the two-particle properties of electron-hole pairs \cite{manzke12}. This was also studied 
for a 2D electron-hole system, considering the exciton self-stabilization mechanism,  
caused by the screening suppression due to the exciton formation \cite{asano14}.
We recently studied a double-layer exciton gas in mean field approximation and found
a transition to a Mott state at high densities \cite{berman15}. Although the destruction
of the superfluid state by strong collisions in the dense exciton gas is plausible,
a mean field approximation is not a very trusted tool to analyze a two-dimensional system.
Therefore, it remains to be shown by an independent and reliable method beyond a mean field approximation that the
long-range superfluid correlations are destroyed in a dense system for all temperatures.

Since an exciton is a strongly bound pair of an electron and a hole, we assume that
these pairs cannot dissociate or transform into a photon by recombination.
This can be justified by a strong Coulomb interaction and a sufficiently short 
time scale that is shorter than the recombination rate. Moreover, we implement a lattice
structure in the layers to allow the excitons to form a commensurate Mott state by
filling each well of the lattice with an exciton. A lattice can be realized by an
electrically charged gate which is periodically structured \cite{gating}.  The gate
can also be used to control the density of excitons via a chemical potential $\mu$. 
Then a pure exciton gas in the periodic potential can be described by the Hamiltonian
\beq
H=-\sum_{rr'}(J_{rr'}+\mu\delta_{rr'})a_r^\dagger a_{r'}+h.c.
\label{hamiltonian00}
\eeq
where the sites $r$, $r'$ are the minima of the potential wells and $J_{rr'}$ is a nearest
neighbor hopping rate:
\[
J_{rr'}=\cases{J & if $r,r'$ are nearest neighbors \cr
0 & otherwise \cr
}
\ .
\]
The lattice structure is characterized by the number of nearest neighbors $c$ (connectivity).
The exciton creation operator $a^\dagger_r$ is composed of the electron creation operator 
$c^\dagger_{e,r}$ and the hole creation operator $c^\dagger_{h,r}$ as
\beq
a^\dagger_r=c^\dagger_{e,r}c^\dagger_{h,r}
\ .
\label{exc0}
\eeq
The form of the exciton operator in Eq. (\ref{exc0}) implies that the excitons can tunnel
freely with the hopping rate $J_{rr'}$ under the restriction that they are composite
particles which obey the Pauli principle. In other words, at most one exciton can occupy
a site of the lattice. Therefore, it is a hardcore boson, similar to the exciton described
by the effective Hamiltonian of Ref. \cite{hanamura77}. 



\section{Conditions for a Mott phase}

Starting from the Hamiltonian (\ref{hamiltonian00}) we consider a grand canonical ensemble of excitons 
with the partition function $Z=Tr e^{-\beta H}$ at temperature $T$ ($\beta=1/k_BT$). The trace $Tr$ is taken
with respect to all exciton states. The exciton correlation function then reads
\beq
\langle e^{\beta tH}a^\dagger_{r}e^{-\beta t H}e^{\beta t'H}a_{r'} e^{-\beta t'H} \rangle
=\frac{1}{Z}Tr e^{-\beta H}e^{\beta tH}a^\dagger_{r}e^{-\beta t H}e^{\beta t'H}a_{r'} e^{-\beta t'H}
\label{corr0}
\eeq
in imaginary time representation with $0\le t-t'\le 1$ \cite{negele}.

The partition function $Z$ of a one-dimensional hard-core Bose gas is given as the determinant of non-interacting
fermions (cf. Ref. \cite{ates05}). The reason for the equivalence of non-interacting fermions and hardcore bosons
in one dimension is the fact that fermions cannot exchange their positions but must obey the repulsive Pauli 
principle. In contrast to a non-interacting Bose gas, this system exhibits two Mott and one an intermediate 
incommensurate phase. Using this example we can expand its free energy $F=-\beta^{-1}\log Z$ in terms of the hopping matrix 
${\bf J}$ as
\beq
F=-\frac{1}{\beta}Tr\log({\bf 1}+e^{\beta\mu}e^{\beta{\bf J}})
=-Tr{\bf J}-\frac{1}{\beta}Tr[\log(1+e^{\beta\mu})]
+\frac{1}{\beta}\sum_{l\ge 1}\frac{(-1)^l}{l} Tr\left[\frac{1}{1+e^{\beta\mu}}(e^{-\beta{\bf J}}-{\bf 1})\right]^l
\ ,
\eeq
provided that $J$ is small in comparison to $\mu$. For $\beta J\gg 1$ (strong quantum
fluctuations) the inequality is valid for $\mu< J$. This example gives us an idea about the convergency of the
hopping expansion in higher dimensions.

For a dilute two-dimensional system we expect a formation of a superfluid state with power law correlations 
$\langle e^{\beta tH}a^\dagger_{r}e^{-\beta t H}e^{\beta t'H}a_{r'} e^{-\beta t'H}\rangle\sim |r-r'|^{-\alpha}$ 
\cite{ziegler02}. On the other hand, 
for sufficiently high density the long-range phase correlations will 
be destroyed by frequent collisions of the excitons. 
This effect leads eventually to an exponentially decaying correlation function 
$\langle e^{\beta tH}a^\dagger_{r}e^{-\beta t H}e^{\beta t'H}a_{r'} e^{-\beta t'H}\rangle\sim \exp(-|r-r'|/\xi)$ with the correlation
length $\xi$. Such a state is either a thermal exciton gas with a fluctuating density at high temperatures or a
Mott state with fluctuating phases but non-fluctuating density at low temperatures. These two states change
from one to the other upon reducing the temperature. If the two phases are connected by a crossover or by a phase
transition is not clear at this point. However, both phases are characterized by a gapped excitation spectrum.
A mean field approximation indicates a crossover \cite{berman15}, where the
Mott phase is characterized by the gap $\Delta_{mf}=\mu-J$ (for $0<J<\mu$) \cite{moseley07}.
Going beyond the mean field approximation and using the dimensionless parameters
\beq
\gamma=\frac{c\beta J}{1+e^{\beta \mu}}\ \ \ {\rm and}\ \
\Delta=\frac{\gamma e^{\beta J}}{1-\gamma}=\frac{c\beta Je^{\beta J}}{1+e^{\beta\mu}-c\beta J}
\label{parameters}
\eeq
we find the following condition for the existence of a Mott phase:
\vskip0.2cm

\no
{\bf Mott correlations:}
{\it For $\mu\ge 0$ and $0\le \Delta <1$ the exciton correlation function decays exponentially as
$0\le\langle e^{\beta tH}a^\dagger_{r}e^{-\beta t H}e^{\beta t'H}a_{r'} e^{-\beta t'H}\rangle\le C_0e^{-|r-r']/\xi}$
with a finite prefactor $C_0$ and a finite decay length $\xi<-1/\log[(1+e^{\beta J})\gamma]$.}
\vs

\no
The derivation of the upper bound is obtained from a hopping expansion, which is given in App. \ref{sect:func_int}.
It should be noted that this expansion consists of two contributions: In terms of Feynman's functional path
integral, the hopping walks of the excitons may or may not cross the time boundaries. The former
is essential for the derivation of the Mott condition $\Delta<1$, whereas the latter requires only the weaker
condition $\gamma<1$.

\begin{figure*}[t]
\begin{center}
\includegraphics[width=9.5cm]{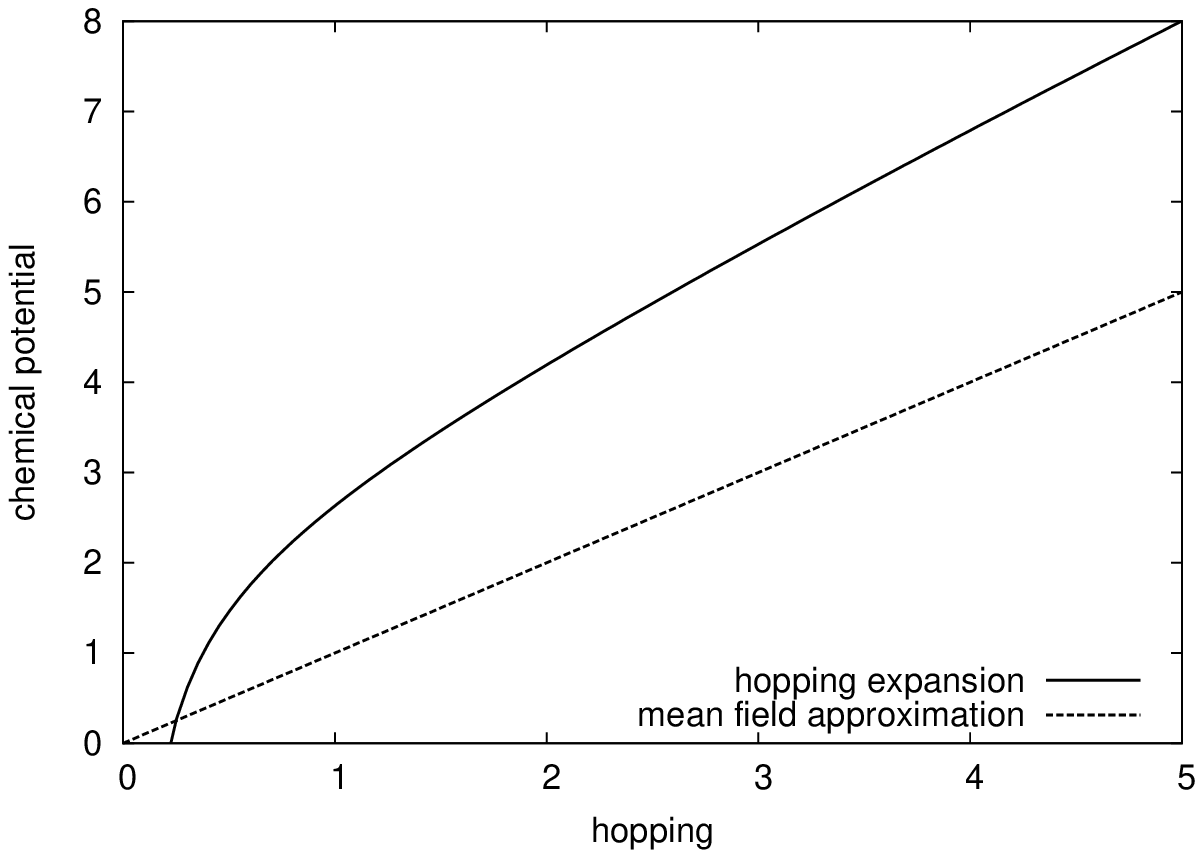}
\caption{
Phase boundary for the dimensionless chemical potential $\beta\mu$ vs. the dimensionless hopping $\beta J$ 
of excitons on a square lattice: A Mott phase with exponentially decaying exciton correlations exists above the solid curve.
The dashed line is the mean field condition $\beta J<\beta\mu$ for a Mott phase. 
}
\label{fig:bound}
\end{center}
\end{figure*}


\section{Discussion}

The interaction in the exciton gas is mediated by the Pauli principle of the constituting fermions (electrons and holes), 
leading to a hardcore interaction of excitons. This repulsive interaction can stabilize a lattice commensurate state, 
where each lattice site is occupied by one exciton. Without hopping (i.e., for $J=0$), such a state exists if the
the lattice density $\rho=1/(1+e^{-\beta\mu})$ is $\rho=1$. This requires $\beta\mu\sim\infty$; i.e., a positive chemical 
potential at a vanishing temperature. We have used in our discussion that the Mott state is characterized by
an exponentially decaying exciton correlations. This characterization is weaker than to fix $n=1$, allowing for
gapped thermal exciton excitations. It leads to the condition $\Delta<1$ or, equivalently,
\beq
c\beta J(1+e^{\beta J})<1+e^{\beta\mu}
\ .
\label{ineq2}
\eeq
This condition reflects the fact that an exciton superfluid state with long-range correlations can always
be destroyed by strong fluctuations for certain values of the hopping rate $\beta J$ and the chemical potential $\beta\mu$.
As an example the phase boundary with $\Delta=1$ on the square lattice (i.e. for $c=4$) is depicted 
in Fig. \ref{fig:bound}. Although the destruction of a superfluid phase seems plausible in the case of suppressed tunneling,
i.e., for $\beta J\ll \beta \mu$, the sufficient condition (\ref{ineq2}) for the appearence of short-range phase correlations
requires a calculation, for instance, in terms of a hopping expansion (cf. App. \ref{sect:func_int}). 

The temperature enters the condition for an exponential
decay only through the normalization of the chemical potential and the hopping rate as $\beta\mu$ and $\beta J$, 
which reflects that either the thermal fluctuations at high
temperatures or the quantum fluctuations at low temperatures are responsible for the exponentially decaying
correlations. As $\beta J$ is a measure for tunneling (i.e., quantum fluctuations), in the case of $\beta J\gg 1$
quantum fluctuations dominate. Then we must provide a sufficient large $\beta\mu$ to obtain Mott correlations:
$\beta\mu>\beta J$. This agrees with the mean field result of Ref. \cite{moseley07}, although the mean field approximation
overestimates the stability of the Mott phase against fluctuations for most values of $\mu$. 
In the high temperature regime with $\beta J\ll 1$, where thermal fluctuations are dominant, 
there is no restriction for $\beta\mu$, which even can vanish. This is also found in the plot of Fig. \ref{fig:bound}.
Thus, in both asymptotic regimes a Mott phase exists. In particular, it is possibles to obtain a Mott phase for $\mu=0$,
where the density is $\rho\approx 0.5$. This situation should be accessible for gated coupled quantum wells. 

In conclusion, as an extension of our previous consideration in Ref. \cite{berman15} for a transition to a Mott state at high densities,
we proved, using a method which goes beyond the mean field approximation, that a bosonic Mott phase exists in an electron-hole bilayer 
through the formation of indirect excitons. In this strongly correlated phase, strong fluctuations destroy the long 
range correlations in the exciton gas at any temperature, either by thermal or by quantum fluctuations.



\appendix

\vskip1cm

\no
{\Large\bf Appendix}

\section{Functional integral representation}
\label{sect:func_int}

To prove the existence of Mott correlations we consider a Grassmann functional integral representation of the partition 
function $Z=Tr e^{-\beta H}$ for space-time variables $x=(r,n)$ with the discrete time $n=1,2,...,M$ \cite{negele}:
\beq
Z=\int_\psi\exp\left(\psi_{1x}\psi_{2x}{\bar\psi}_{2x}{\bar\psi}_{1x}
+\psi_{1x}\psi_{2x}v_{xx'}{\bar\psi}_{2x'}{\bar\psi}_{1x'}\right) 
, \ \ \ 
v_{rn,r'n'}=\cases{
1+\frac{\beta}{M}\mu & for $r'=r$, $n'=n+1$ \cr
\frac{\beta}{M}J_{rr'} & for $r'\ne r$, $n'=n+1$ \cr
0 & otherwise \cr
}
\label{ferm_rep2}
\eeq
with the Grassmann integration $\int_\psi$ \cite{negele}. At the end we take the limit $M\to\infty$. 
It should be noticed that the Grassmann variables
$\psi_x$ are anti-periodic in the time direction. 

The correlation function of Eq. (\ref{corr0}) reads in terms of the functional integral
\beq
\langle {\bar\psi}_{2y}{\bar\psi}_{1y}\psi_{1y'}\psi_{2y'}\rangle
=\frac{1}{Z}\int_\psi{\bar\psi}_{2y}{\bar\psi}_{1y}\psi_{1y'}\psi_{2y'}\exp\left(\psi_{1x}\psi_{2x}{\bar\psi}_{2x}{\bar\psi}_{1x}
+\psi_{1x}\psi_{2x}v_{xx'}{\bar\psi}_{2x'}{\bar\psi}_{1x'}\right) 
\label{corr2}
\eeq
with $y=(r,Mt)$ and $y'=(r',Mt')$. It is convenient to introduce the generating functional
\beq
Z(\{\alpha_{yy'}\})
=\int_\psi\exp\left(\psi_{1x}\psi_{2x}{\bar\psi}_{2x}{\bar\psi}_{1x}
+\psi_{1x}\psi_{2x}(v+\alpha)_{xx'}{\bar\psi}_{2x'}{\bar\psi}_{1x'}\right) 
\eeq
and take the derivative of $\log(Z(\{\alpha_{yy'}\})$. This, for instance, provides
for the correlation function
\[
\langle {\bar\psi}_{2y}{\bar\psi}_{1y}\psi_{1y'}\psi_{2y'}\rangle
=\frac{\partial}{\partial \alpha_{y'y}}\log(Z(\{\alpha_{yy'}\})\Big|_{\alpha=0}
\ .
\]

\begin{figure*}[t]
\psfrag{r}{$r$}
\psfrag{r'}{$r'$}
\psfrag{m}{$m$}
\psfrag{m'}{$m'$}
\begin{center}
\includegraphics[width=8cm]{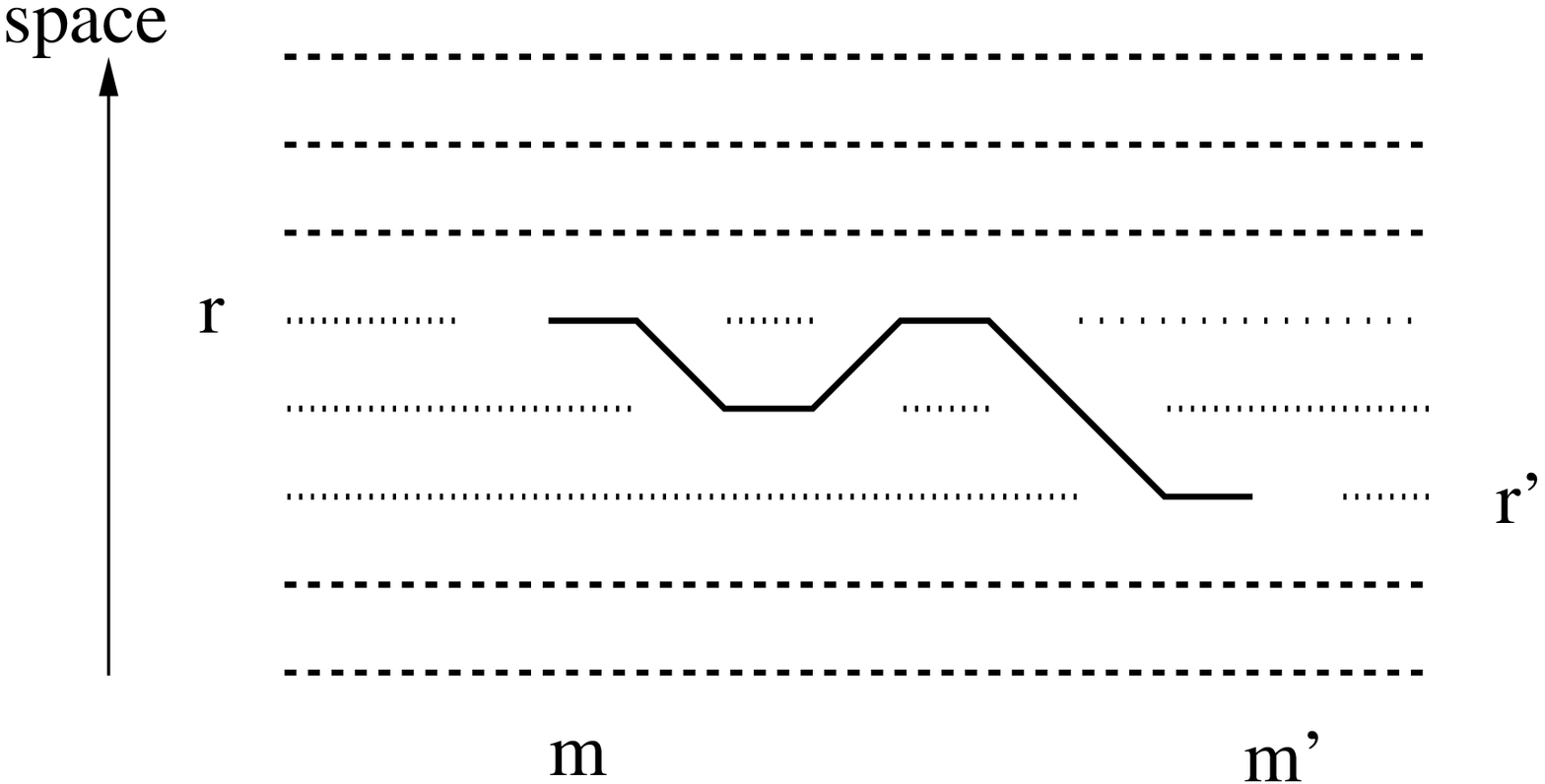}
\includegraphics[width=8cm]{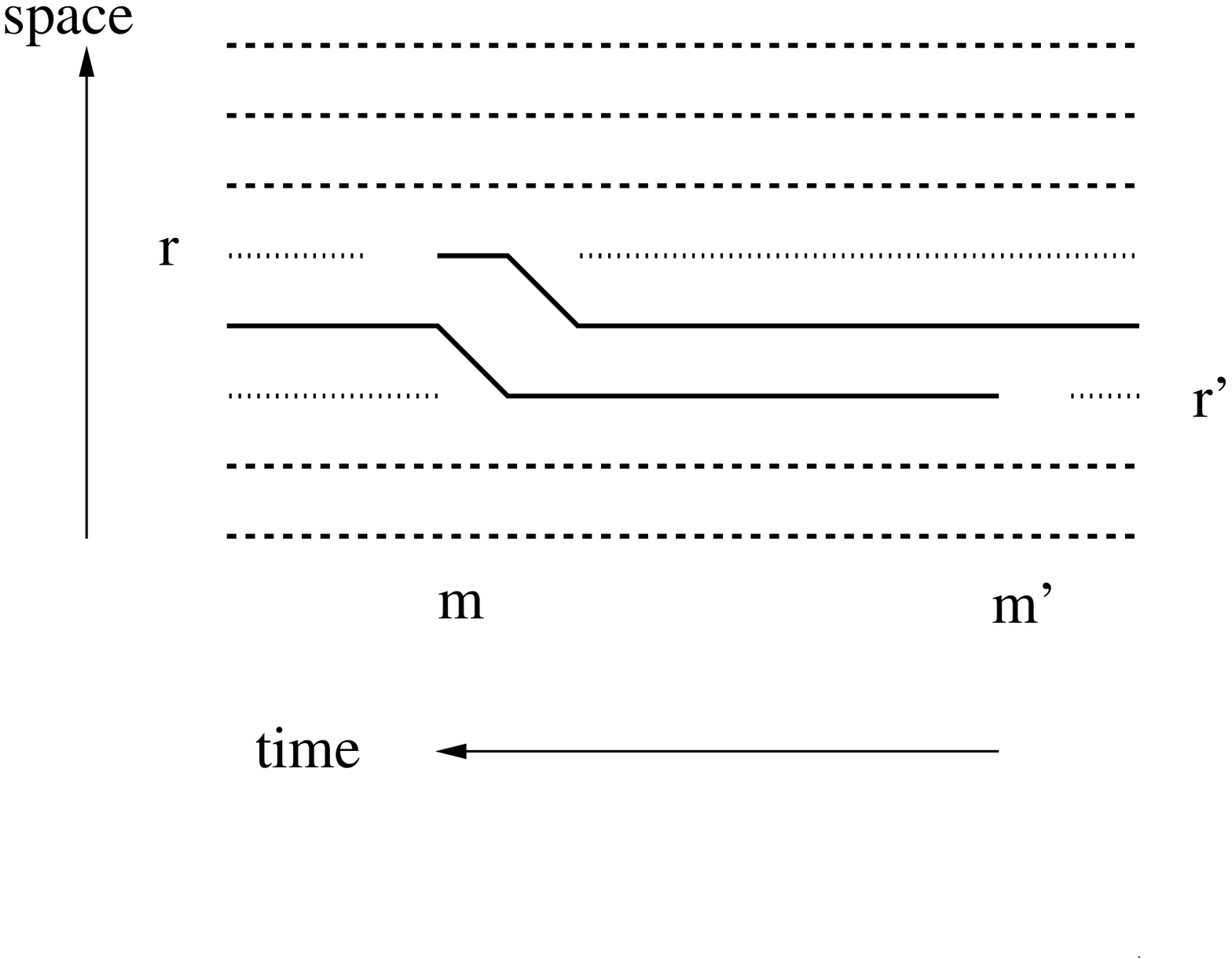}
\caption{
Walk from $y=(r,m)$ to $y'=(r',m')$. The thick line represents the connecting walk, each dashed line
has the weight $1+e^{\beta\mu}$ and the dotted line weight 1. There is either a walk that connents
$y$ and $y'$ directly (upper example) or through the periodic boundaries in time direction (lower example).
The weight of the walk is estimated in (\ref{est1}).}
\label{fig:walk}
\end{center}
\end{figure*}

\subsection{Mott phase}
\label{sect:mott}

The Grassmann integral (\ref{ferm_rep2}) can be directly calculated if $J_{rr'}=0$ (i.e., in the absence of exciton hopping), 
giving $Z_M=[1+(1+\beta\mu/M)^M]^N$ for a lattice with $N$ sites. For $M\to\infty$ we obtain 
$\lim_{M\to\infty}Z_M= (1+e^{\beta\mu})^N$. In this case the exciton density is
$\rho=1/(1+e^{-\beta\mu})$. This suggests that we consider the hopping term
\beq
\psi_{1x}\psi_{2x}\frac{\beta}{M}w_{xx'}{\bar\psi}_{2x'}{\bar\psi}_{1x'}
\ \ \ (w_{xx'}=J_{rr'}\delta_{n',n+1})
\eeq
as a perturbation and apply the linked cluster approach for $\log Z(\{\alpha_{xx'}\})$, which
leads to a connected walk with end points $y$ and $y'$ \cite{glimm} (cf. Fig. \ref{fig:walk}). 
In other words, we expand in (\ref{corr2})
\[
{\bar\psi}_{2y}{\bar\psi}_{1y}\psi_{1y'}\psi_{2y'}\sum_W\prod_{(x,x')\in W}
\exp(\psi_{1x}\psi_{2x}v_{xx'}{\bar\psi}_{2x'}{\bar\psi}_{1x'})
\]
\beq
={\bar\psi}_{2y}{\bar\psi}_{1y}\psi_{1y'}\psi_{2y'}\sum_W\prod_{(x,x')\in W}
(1+\psi_{1x}\psi_{2x}v_{xx'}{\bar\psi}_{2x'}{\bar\psi}_{1x'})
\label{exp2}
\eeq
along the walk $W$ between $y$ and $y'$. Here it should be noticed that $W$ can either connect $y$ and $y'$
directly (upper example in Fig. \ref{fig:walk}) or by using the Grassmann anti-periodic boundaries in time
(lower example in Fig. \ref{fig:walk}). The latter case will be called a walk with boundary crossing.

After the Grassmann integration expression (\ref{exp2}) gives 
for walks with $k$ boundary crossings a sequence of $m-m'+Mk$ factors $v_{xx'}$:
\beq
\langle {\bar\psi}_{2y}{\bar\psi}_{1y}\psi_{1y'}\psi_{2y'}\rangle
=\sum_{k\ge 0}\sum_{x_1,...,x_{m-m'+Mk}\in W}(1+e^{\beta\mu})^{-n_v}v_{yx_1}v_{x_1x_2}\cdots v_{x_{m-m'+Mk}y'}
\ ,
\eeq
where $n_v$ is the number of visited spatial sites. 
The product of the $v$ terms for a fixed walk $W$ from $z=(r,m)$ to $z'=(r',m')$ with $0\le m'\le m\le M$
and without boundary crossing is estimated as
\beq
0\le v_{zx_1}v_{x_1x_2}\cdots v_{x_{m-m'}z'}\le \cases{
\left(1+\frac{\beta}{M}\mu\right)^{m-m'} & $n=1$ \cr
\left[1+\frac{\beta}{M}(\mu+J)\right]^{m-m'-n_v+1}(\beta J)^{n_v-1} & $n\ge 2$ \cr
}
\ .
\eeq
Now we use the time variables $t=m/M$, $t'=m'/M$ and take the limit $M\to \infty$ to obtain
\beq
0\le (1+e^{\beta\mu})^{-n_v}v_{zx_1}v_{x_1x_2}\cdots v_{x_{m-m'}z'}\le 
\cases{
e^{\beta\mu(t-t')}/(1+e^{\beta\mu}) & $n_v=1$ \cr
[\beta J/(1+e^{\beta\mu})]^{n_v-1}e^{\beta(\mu+J)(t-t')}/(1+e^{\beta\mu}) & $n_v\ge 2$ \cr
}
\ .
\label{est1}
\eeq
Moreover, we have
\beq
e^{\beta(\mu+J)(t-t')}/(1+e^{\beta\mu})\le e^{\beta J(t-t')}\le e^{\beta J}
\ ,
\eeq
since $0\le t-t'\le 1$. As a time independent upper bound we can choose $t=1$, $t'=0$.


Finally, we must sum over all possible walks $W$. There are $c$ (number of nearest neigbors) choices for each hop
of an exciton. With $k-1$ boundary crossings this implies for the correlation function
\beq 
\langle {\bar\psi}_{2y}{\bar\psi}_{1y}\psi_{1y'}\psi_{2y'}\rangle
\le\sum_{n\ge 1}a_n\Theta_{n}(|r-r'|)
+\sum_{k\ge2}\sum_{n_1,...,n_k\ge 2}a_{n_1}\cdots a_{n_k}\Theta_{n_1,...,n_k}(|r-r'|)
\ ,
\label{ineq3}
\eeq
where
\[
\Theta_{n_1,...,n_k}(|r-r'|)=\cases{
1 & for $n_1+\cdots +n_k -k\ge |r-r'|+1$\cr
0 & otherwise\cr
}
\ , \ \ \ 
a_n=\cases{
1 & $n=1$ \cr
e^{\beta J}\gamma^{n-1} & $n\ge 2$\cr
}
\]
and $\gamma$ is defined in Eq. (\ref{parameters}).
$\Theta_{n_1,...,n_k}(|r-r'|)$ enforces that at least $|r-r'|+1$ sites are visited to connect the sites $r$ and $r'$.
Next we perform the summation in Eq. (\ref{ineq3}) without the factor $\Theta_{n_1,...,n_k}(|r-r'|)$ and obtain
\beq
\sum_{n\ge 1}a_n+\sum_{k\ge2}\sum_{n_1,...,n_k\ge 2}a_{n_1}\cdots a_{n_k}
=1+e^{\beta J}\sum_{n\ge2}\gamma^{n-1}+\sum_{k\ge2}e^{k\beta J}\left(\sum_{n\ge2}\gamma^{n-1}\right)^k
=\frac{1+(e^{\beta J}-1)\gamma}{1-\gamma}+\frac{\Delta^2}{1-\Delta}
\ ,
\label{summation0}
\eeq
provided that $\Delta = \gamma e^{\beta J}/(1-\gamma)<1$. This condition can also be written as
\beq
c\beta J(1+e^{\beta J})<1+e^{\beta\mu}
\ .
\eeq
Thus, the contribution of the summation of crossings is a factor $1+e^{\beta J}$.
Finally, we must include the factor $\Theta_{n_1,...,n_m}(|r-r'|)$ in the summation to obtain the upper
bound in (\ref{ineq3}). This is equivalent of taking in (\ref{summation0}) only terms into account 
with powers $\gamma^{|r-r'|+1}$ and higher. The first term on the right-hand side of Eq. (\ref{summation0})
is a geometric series in powers of $\gamma$ and the second term a geometric series in powers of
$\gamma(1+e^{\beta J})$, since
\[
\frac{\Delta}{1-\Delta}=\frac{\gamma e^{\beta J}}{1-\gamma(1+e^{\beta J})}
=\gamma e^{\beta J}\sum_{l\ge 0}[\gamma(1+e^{\beta J})]^l
\ .
\] 
Thus, the second term gives the leading contribution in Eq. (\ref{summation0}), which allows us to extract
\beq
\langle {\bar\psi}_{2y}{\bar\psi}_{1y}\psi_{1y'}\psi_{2y'}\rangle
\le C_0\left(\frac{c\beta J(1+e^{\beta J})}{1+e^{\beta\mu}}\right)^{|r'-r|}
\eeq
as an upper bound for the correlation function (\ref{ineq3}).

\end{document}